# A mechanism of Individualistic Indirect Reciprocity with internal and external dynamics

Mario Ignacio González Silva, Ricardo Armando González Silva[*], Héctor Alfonso Juárez López y Antonio Aguilera Ontiveros
Universidad de Guadalajara, Centro Universitario de los Lagos.
Enrique Díaz de León 1144, Paseos de La Montaña, Lagos de Moreno, Jal.
47460 / Lagos de Moreno Jalisco / México
Colegio de San Luis
Parque de Macul #155, Fracc. Colinas del Parque,
San Luis Potosi, S.L.P., Mx., C.P. 78294

**Abstract**

The cooperation mechanism of indirect reciprocity has been studied by making multiple variations of its parts. This research proposes a new variant of Nowak and Sigmund model, focused on agents' attitude; it is called Individualistic Indirect Reciprocity. In our model, an agent reinforces its strategy to the extent to which it makes a profit. We also include conditions related to the environment, visibility of agents, cooperation demand, and the attitude of an agent to maintain his cooperation strategy. Using Agent-Based Model and a Data Science method, we show on simulation results that the discriminatory stance of the agents prevails in most cases. In general, cooperators only appear in conditions with low visibility of reputation and a high degree of cooperation demand. The results also show that when the reputation of others is unknown, with a high obstinacy and high cooperation demand, a heterogeneous society is obtained. The simulations show a wide diversity of scenarios, centralized, polarized, and mixed societies.

## Introduction

Why is cooperation between humans so tricky? What is required for cooperation to exist? How to establish cooperation in a world of selfish people (without a central authority)? or When should a person cooperate, and when should he be selfish (in frequent interaction)? Cooperation has been identified as one of the paths for personal and entities development, say for towns, cities, countries, etc. The most crucial aspect of evolution is the ability to generate cooperation in a competitive world (Nowak, 2006). A simple definition of cooperation is that one individual pays a cost for another to receive a benefit. Cost and benefit are measured in terms of reproductive success (Rand & Nowak, 2013).

---

[*] Corresponding Author

The evolution of human cooperation had been explained by five main mechanisms: direct reciprocity, indirect reciprocity, spatial selection, multi-level selection, and relationship selection. These main regularities of interaction are called mechanisms, which are necessary for the evolution of cooperation, and they are very different from behaviors that require evolutionary explanation (such as strong reciprocity, upward reciprocity, and parish altruism) (Rand & Nowak, 2013).

The indirect reciprocity operates if there are repeated encounters within a population and third parties observe or know about some of these encounters. Information about these meetings can be spread through communication, affecting the reputation of the participants. Individuals can adopt conditional strategies that base their decision on the recipient's reputation. Its main variants are rewards (why altruism spreads), punishment (why rules spread), and deception (why cheating spreads). The cost-benefit calculus for cooperation is not always carried out deliberately or consciously; the return is expected from someone other than the recipient of the benefit. Indirect reciprocity develops because interactions are repeated or flow among a society's members and because information about subsequent interactions can be gleaned from observing the reciprocal interactions of others. Indirect reciprocity involves reputation and status and results in everyone in a social group continually being assessed and reassessed by interactants, past, and potential, based on their interactions with others. Indirect reciprocity presupposes rather sophisticated players and therefore is likely to be affected by anticipation, planning, deception, and manipulation (Alexander, 1987).

One of the simulation research guidelines for the indirect reciprocity mechanism based on the image score and k-strategy cooperation (where k-strategy is an integer value between -5 and 6) was proposed by Nowak and Sigmund (Nowak & Sigmund, 1998a). The main investigations that contextualize this research and make up this type of modeling are the following.

Nowak y Sigmund (Nowak & Sigmund, 1998a) made the first theoretical and simulated work based on Alexander's statement of indirect reciprocity. They show that individual selection can nevertheless favor cooperative strategies directed towards recipients that have helped others in the past. Later they show analytically in (Nowak & Sigmund, 1998b) that discriminating altruism can be resistant to invasion by defectors. Indiscriminate altruists can invade by random drift, however, setting up a complex dynamical system.

Lotem et. al in (Lotem, Fishman, & Stone, 1999) make a realistic experiment of indirect reciprocity mechanism, allowing individuals to carry D phenotype defectors with a non-heritable phenotype strategy k=+7. They conclude a paradoxical effect whereby phenotype defectors stabilize discriminate altruism can be explained analytically. Another experimental research designed to distinguish between the two proposed mechanisms of indirect reciprocity: discriminator image scoring and standing strategies, was done in (Manfred Milinski, Semmann, Bakker, & Krambeck, 2001). They studied 23 groups comprising seven players each, determine that standing strategies demand too much work and a large amount of second- if not third- and fourth-order information about the history of the social interactions.

A research on the evolution of indirect reciprocity based on image scoring is the strategy of aiming for "good standing"; it was studied in (Leimar & Hammerstein, 2001). They show that it has superior properties, it can be an evolutionarily stable strategy, and, in some cases, it tends to outperform the image score. Another model with a framework for the evolution of indirect reciprocity by social information is proposed in (Mohtashemi & Mui, 2003). Its information is selectively retrieved from and propagated through dynamically evolving networks of friends and acquaintances. They show analytically that for indirect reciprocity to be evolutionarily stable, the differential probability of trusting and helping a reputable individual over a disreputable individual, at a point in time, must exceed the cost-to-benefit ratio of the altruistic act.

Another paper that studies indirect reciprocity with experimental evidence is in (Seinen & Schram, 2006). In their experiments, indirect reciprocity is largely based on norms about how often the recipient should have helped others in the past. They show that these norms develop similarly within groups of interacting subjects but distinctly across groups, which leads to the emergence of group norms.

The purpose of this research is to model and simulate the behavior of agents under the individualistic indirect reciprocity mechanism adding external factors such as the visibility of the agents (or ability to see the reputation of others), obstinacy when changing strategies, and the cooperation demand for compelling them to interact cooperatively. We want to characterize when cooperation is established or not according to the different values of the simulation parameters. In this paper, we present an application of the Agent-Based Model methodology, developed in NetLogo and analyzed with Python with three parameters without a social network. It extends the work of (González Silva, González Silva, & Juárez López, 2020), refining the parameters value, Hierarchical Cluster Method analysis, and its simulation results' characterization.

The article is structured into six sections. The first section is a literature review, where we present the articles that support our research. The second section is about Indirect reciprocity and individualism indirect reciprocity mechanisms; we show the theories that support it and the foundations of individualism. In the Materials and Methods section, we present all the elements used for this research, as well as the experiment design and the simulation process. In the Simulation Results section, we show and systematically describe the results of the experiments. We analyze these results in the Discussions section. Finally, the conclusion section has the fundamental observations of the research, and we add the ODD protocol to our model in the appendix.

## Indirect reciprocity mechanism and individualism

Alexander shows that the essence of moral systems seems to lie in patterns of indirect reciprocity (Alexander, 1979, 1981). He establishes that Indirect reciprocity is what happens when direct reciprocity occurs in the presence of an interested audience in the interaction of cooperation (see Figure 1).

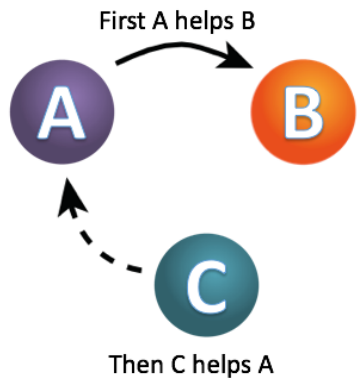

**Figure 1. Indirect reciprocity cooperation mechanism** (Nowak & Sigmund, 2005)

The mechanism of indirect reciprocity has been extensively studied. The investigations about the mechanism of indirect reciprocity have analyzed each of its elements with its variants, they have determined new types, but the most analyzed is the downstream type (Okada, 2020), it has been studied under various conditions, for example, by adding social norms (Ohtsuki & Iwasa, 2006). Also, it has been studied with mathematical theories such as game theory, computational methods, social, biological (see references of (Rand & Nowak, 2013) or (Nowak & Sigmund, 2005)) and even experimentally (as presented in the literature review section).

Can altruism emerge in an individualistic population? Under what conditions or norms is there altruism in an individualistic society? The concept of individualism has evolved since it was coined and has had various connotations, and has basically been studied in social, psychological, philosophical, and economic contexts (Lukes, 2006). Oxford dictionary defines individualism "as the quality of being different from other people and doing things in its way, i.e., the behavior of someone who does things in their way without worrying about what other people think or do."

Without going into philosophical conceptions, we can consider individualism as a basic criterion for deciding or acting for people; it is the criterion of prioritizing for one's benefit. Individualism is a characteristic of society; situations always arise in which people are not willing to do something without first having their benefit.

In the Indirect Reciprocity Mechanism simulation models with k-strategy, they consider that the agent changes strategy based on the inheritance or evolution of the next generation of agents. We propose that the agent change its strategy based on a criterion of "individualistic" (figure 2) with levels of flexibility (stubbornness). We want to analyze what happens with cooperation in the MCRI.

How to characterize the flexibility of acceptance of new ideas or impositions? How to identify some element of the agents to give a sense of flexibility's criteria of change of ideas?

Oxford dictionary defines *obstinacy* as the attitude of somebody who refuses to change their opinions, way of behaving, Etc., when other people try to persuade them to; behavior that shows this. To solve these questions in our model, agents can reformulate their strategy at the first good/bad experience or after several similar experiences. In general, to have harmony in societies, we deal with laws or precepts: explicit, implicit, local, global, Etc. In this sense, in the model, we add a rule to compromise the cooperation of the agents.

## Individualistic Indirect reciprocity mechanism (IIRM)

The individualistic indirect reciprocity mechanism is like Nowak's style. The fundamental difference is how the strategy is updated; we describe it below.

For a given agent, if its strategy has a negative value, we will call it "a cooperator". If its strategy has a positive value, we will call it "a defector". If the value of its strategy is zero, we will call it "a discriminator". We define the Individualistic Indirect Reciprocity Cooperation Mechanism (MC-RII) as the act in which an agent "reinforces" his strategy based on the instantaneous gains of past interactions; that is, if he is cooperative and someone cooperates with him, he will increase his cooperation index. Similarly, if it is a defector and someone cooperates with him, its index will decrease (Gonzalez-Silva, 2018). This is represented in Figure 2.

Each mark represents a situation where the strategy of each agent can be; as in case 2a) If an agent has the strategy $k = -2$ and someone cooperates with him, he will move his strategy to $-3$, but if the other does not cooperate with him, he will move towards $-1$. For $k = 3$, and someone cooperates with him, the agent will reinforce his strategy by going to the right to $k = 4$, however, if the agent does not receive cooperation, he will move to the left to $k = 2$. 2b) shows when the agents are discriminators, they become cooperators if they cooperate and defectors if they do not. Figure 2c) shows when the agents have extreme strategies $k = -5$ and $k = 6$, the former are unconditional cooperators, and the others are extreme defectors. If an agent has the strategy $k = -5$, if someone cooperates with him, they will stay in their strategy, but if not, he will move to the left.

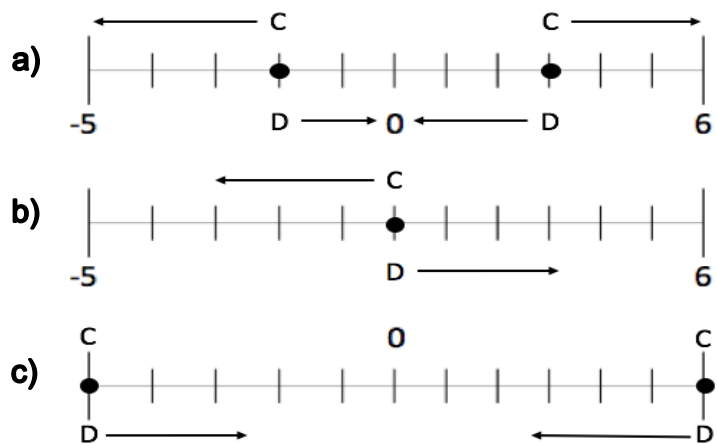

Figure 2. Indirect individualistic reciprocity with reinforcement

## Simulation Process of IIRM

Our simulation process starts with a model description; we use the ODD (see appendix). The ODD protocol is a standard for describing simulation models to help promote rigorous formulation (Grimm et al., 2020). To describe the elements of the model, we use Agent-based Models (ABM) methods, it is a class of computational models used to simulate the actions and interactions of autonomous agents and in order to assess their effects on the system as a whole (Paranjape, Wang, & Gill, 2018). In order to develop the simulations, we use NetLogo (Wilensky, 1999) (https://ccl.northwestern.edu/netlogo/) and Python software to analyze simulation results with the tool Hierarchical Cluster Analysis to see the connection between the objects inside the cluster, and the agglomerations were expressed as a dendrogram (King, 2015). Finally, we use the Convex hull for Scientific Visualization (Rockafellar, 1990).

In our model, we establish with a certain probability the known (for all other agents) of the image score value of each individual (we call to this probability the "visibility"). The agents are randomly chosen at each iteration and those that interact update their strategy, reputation, and payments.

The population is composed of $n$ agents. Each agent $i$ has a strategy $k_i$ (that takes integer values in the range -5 to 6 inclusive), a reputation $s_i$ (that takes integer values in the range -5 to 5 inclusive), and a payment function $p_i$. At each iteration, there is a fixed number $m$ of interactions of pairs of agents. Within each pair, one agent is randomly chosen to be the "donor" (denoted by $i$) and the other is the "recipient" (denoted by $j$). With probability V, the donor $i$ knows the reputation of agent $j$, he cooperates based on the cooperation criterion $k_i < s_j$, if not, the cooperation demand criterion is applied.

When donor $i$ cooperates, the value of its reputation is increased in one unit, and if he does not cooperate, its reputation is decreased by one unit. When cooperation happens, the donor pays a cost $c$, and the recipient gets a profit $b$; there is no reward in the absence of cooperation. The donor's strategy and the recipient's reputation remain unchanged. However, the recipient's strategy changes based on the obstinacy parameter.

The model has three controllable parameters: visibility, with values from 0 to 1 with step change of 0.1; obstinacy, with values from 0 to 6 with change step of 1; and cooperation demand that takes values 0,1, 2, 3, 4, 5, and 6. We briefly describe each one of them.

Visibility parameter ($V$), with values from 0 to 1. In the process of cooperation, a donor agent $i$ will "see" (or know) the reputation of the recipient agent $j$ with probability $V$. If he sees the reputation, he applies the cooperation criterion $k_i < s_j$. If a donor cannot see the recipient's reputation, he will suppose that the recipient has a "good reputation" and will go to continue the cooperation process based on the value of the cooperation demand parameter.

Obstinacy parameter ($O$), with values from 0 to 6. In the IIRM model, an agent is going to reinforce its strategy when he obtains profits and changes on the opposite side (cooperator/defector) if he does not get rewards. This behavior is controlled by the obstinacy parameter. If the obstinacy parameter is $n$, then an agent will reinforce (or change) its strategy only after $n$ consecutive times that he was the receptor of cooperation (or he was not). An obstinacy parameter of 0 means that the agent will reinforce its strategy at each interaction.

Cooperation demand parameter ($CD$), with integer values from 0 to 6. This parameter forces the agents to cooperate. If the cooperation demand parameter has value $l$, the cooperation is forced for those donor agents that we're unable to see the recipient's reputation and hold $k_i \leq l$.

When the values of k vary from -1 to 1 the agents are considered discriminators, if $k \leq -2$ they will be cooperators and $k \geq 2$ delators. Furthermore, we will consider the feature space of the histograms the space generated by their mean, variance, and skewness parameters.

# Experimental Design

The following initial conditions are common to all experiments. A population of $n = 100$ agents is set up. The values of strategy $k$ are randomly drawn with uniform distribution. The value of the reputation $s$ of each agent, as well as the value of its payment function, are set to 0. The benefit ($b$) and cost ($c$) parameters are set to $b = 1, c = 0.1$ (to avoid negative profits, we add 0.1 at the beginning of each interaction).

For each experiment, the values of the main parameters are set: visibility (0 to 1), obstinacy (0 to 6), and cooperation demand (can be 0,1, 2, …, 6). At each experiment, once all the parameters have been set up, we perform 201 iterations (tics or time steps). At each iteration or time step, we randomly select m pairs of agents to interact (in each pair one of the agents is selected to be the donor and the other to be the recipient). Thus, each agent has an average of $2m/n$ interactions.

Once that a pair of agents have interacted, their strategies, payments, and reputations are updated asynchronously. The recipient reinforces its strategy based on if he received cooperation (or not) and the obstinacy parameter, adds $b$ to its payments (if he received cooperation) and its reputation is unchanged. The donor leaves unchanged its strategy, increases its reputation by 1 in the case of have been cooperative (or subtracts 1 if he refused to cooperate) and subtracts $c$ if he cooperates ($c$ is the cost of cooperating).

Taking into account the values of the three main parameters (visibility V, obstinacy O, and cooperation demand CD), we have a total of 539 possible combinations.

## Simulations' Hierarchical clustering analysis

Each simulation generated values for strategy, reputation, and playouts, for every iteration of time t. The final simulations' results of the k strategy are structured in a histogram with 12 values distributed from -5 to 6. We use three of their graphic characteristics to analyze these histograms: mean, standard deviation, and skewness. We call this group of parameters a characteristic space.

With the characteristics space, a new database was generated. The first column corresponds to the simulation number, the second the mean value, the third contains the standard deviation, and the fourth the asymmetry value.

We use hierarchical clustering analysis to group the points of the characteristics space by similarity and generate a dendrogram with the Ward method. This analysis generated 8 clusters by partitioning at height 2 (this means that the centroids of the clusters are at a distance less than 2 in the characteristics space). Although previously, the histograms with zero standard deviation were isolated, generating 9 clusters.

We proceeded to visualize how the different groupings are distributed using the model's parameters, using a three-dimensional space; in the axes, we place visibility, obstinacy, and the cooperation demand. Thus, we finally visualize 9 clusters in space and generate the convex envelope to join the set of points, to have a more defined projection of the clusters, and to each one of them attach us representative histograms.

## Simulation Results

To analyze the cooperation dynamics, we will focus on the final frequency distribution of strategies, and the behavior of the temporal average of payments, and the temporal average of reputations for the stability of the simulations. The final simulation results were grouped into 9 cases.

Figure 3 shows selected insights of the temporal evolution of the experiment for $V = 0, O = 0$ y $CD = 1$. Figure 3a) shows initial conditions of the population ($t = 0$), a heterogeneous society. 3b) shows the temporal state at $t = 25$. 3c) display the temporal state for t = 100. The image 3d) exhibit the final estate of the simulation. In this case, we observe that the society is distributed to the side of defectors, decreasing in number as the agents' strategy tends to 6.

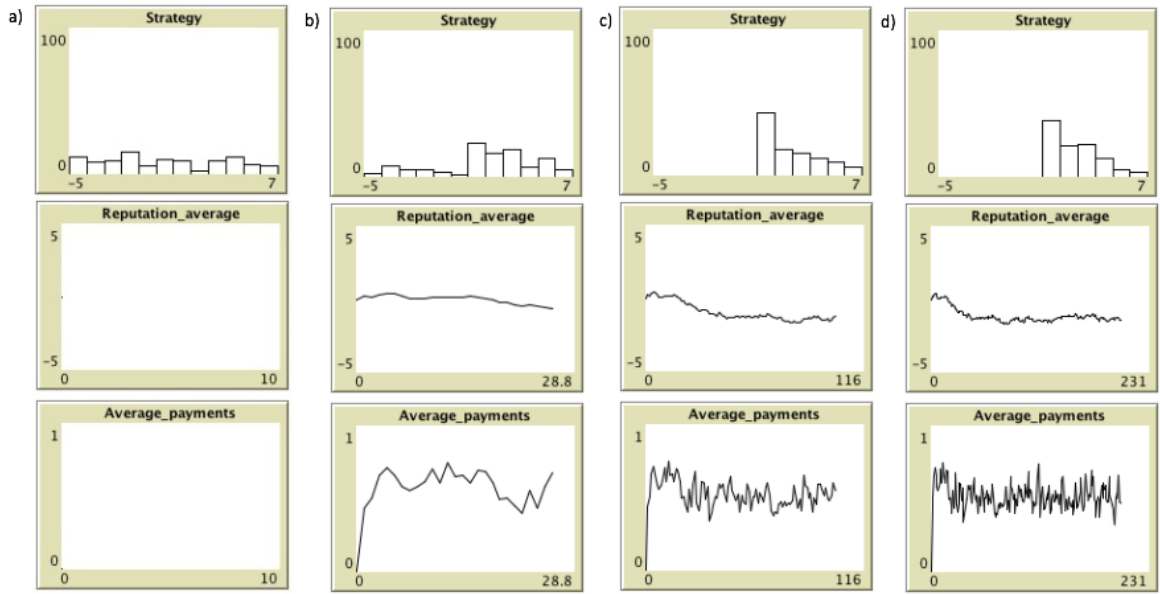

**Figure 3. Temporal evolution t=0, t=25, t=100 and t=201.**

Figure 4 shows two representative histograms for clusters 1 through 8 and one for cluster 9, which have very similar values in feature space, but with some difference in skewness. It also shows the convex-hull of each cluster determined by the dendrograms at height 2 in the parameter space.

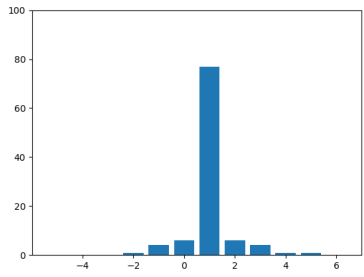 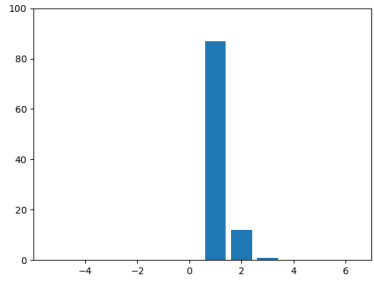 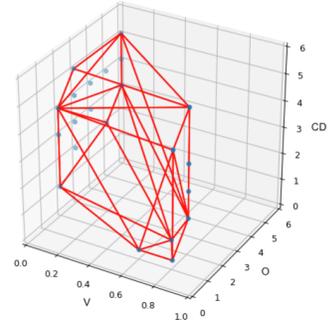

Figure 4a) Cluster 1.

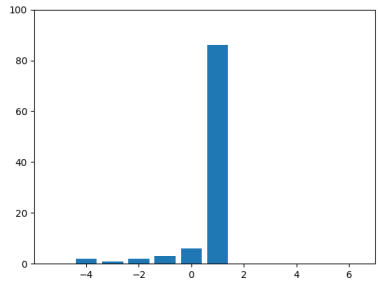 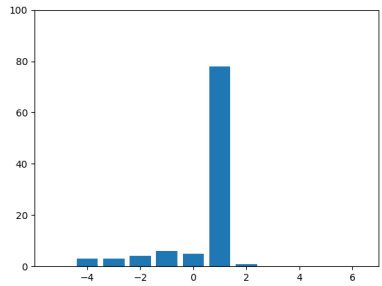 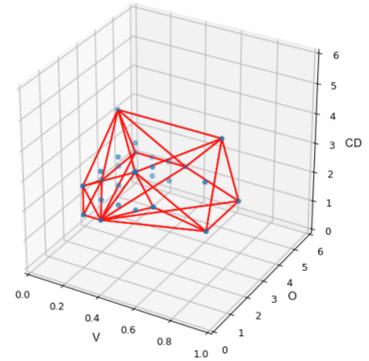

Figure 4b) Cluster 2.

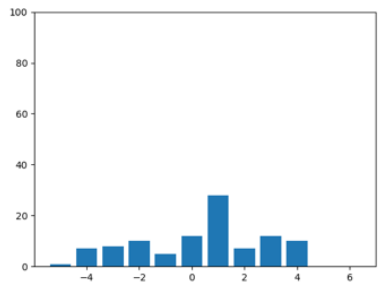 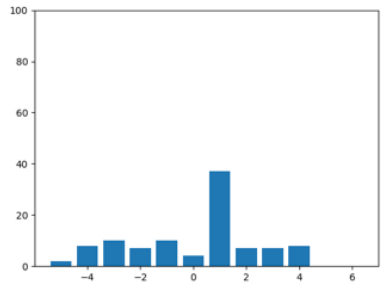 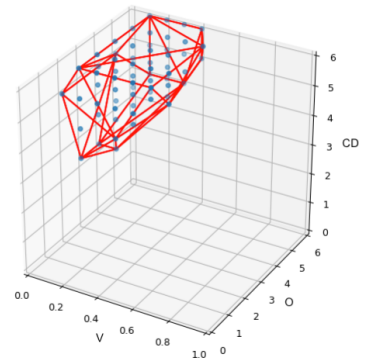

Figure 4c) Cluster 3.

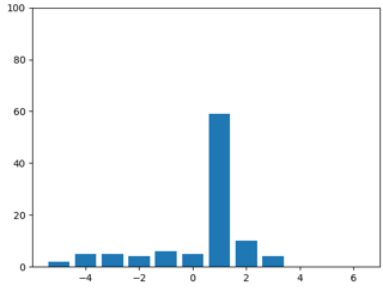 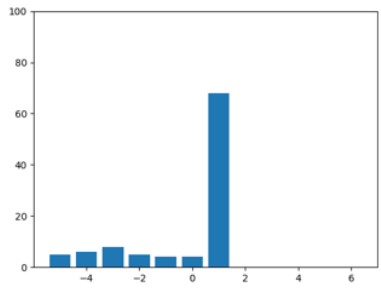 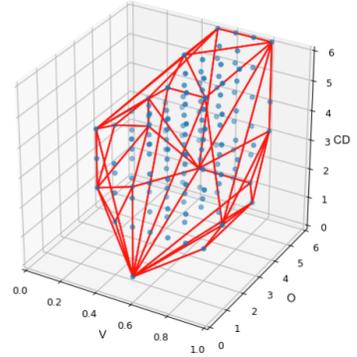

Figure 4d) Cluster 4.

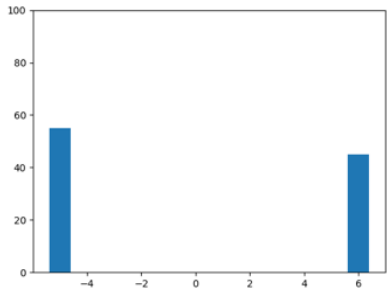 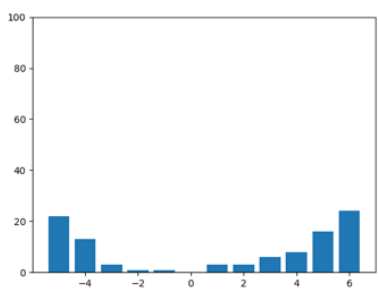 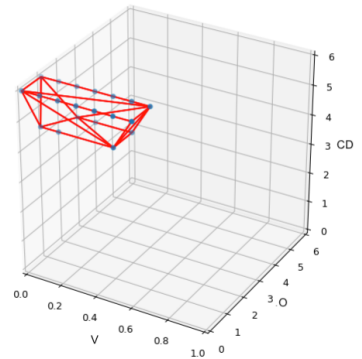

Figure 4e) Cluster 5.

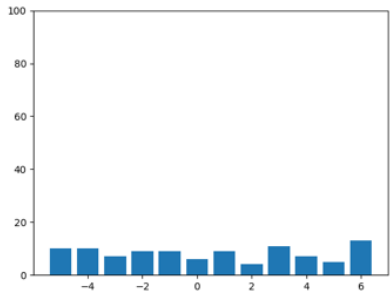 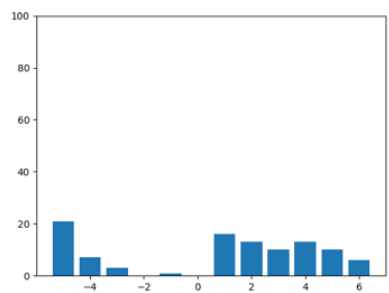 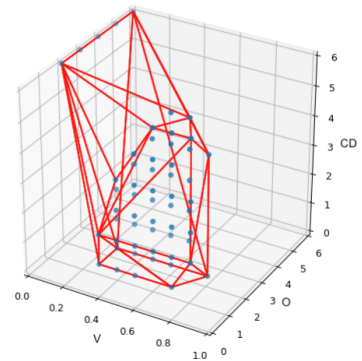

Figure 4f) Cluster 6.

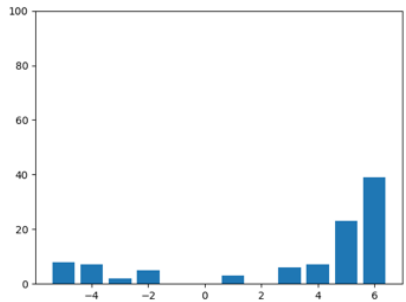 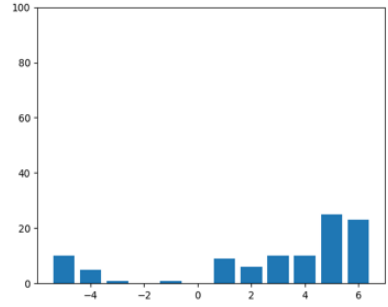 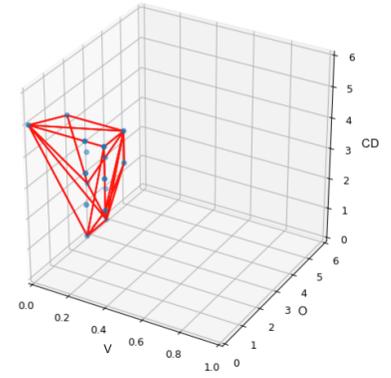

Figure 4g) Cluster 7.

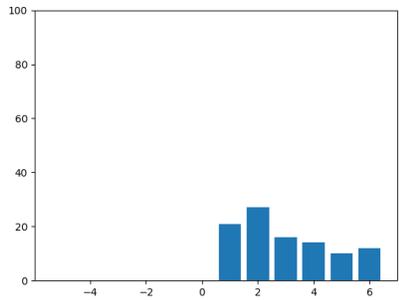 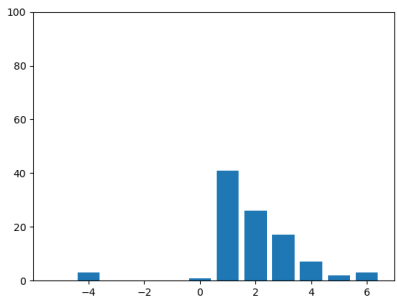 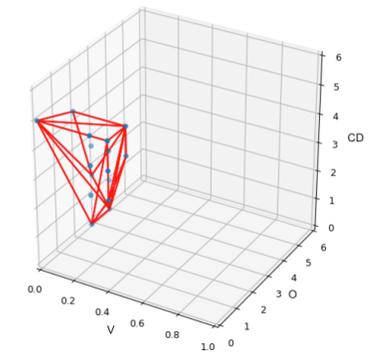

Figure 4h) Cluster 8.

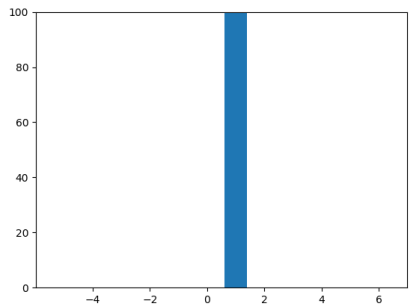 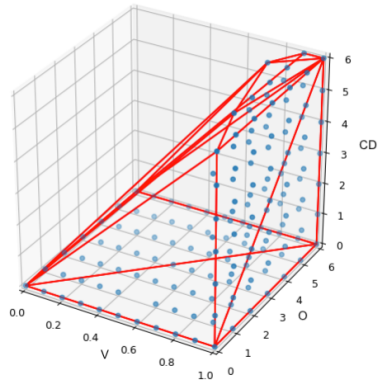

Figure 4i) Cluster 9.

**Numerical description by cluster.**

Cluster 1: (22 histograms) We observe that this cluster has two groups, the first group with zero visibility, obstinacy with a value from 2 to 6, and cooperation requirement from 1 to 5; in this, the proportions of the strategy are at 90% in bin number 1, the other 10% are grouped around. The second group with visibility from 0.7 to 0.9 obstinacy at 0 and 1, cooperation requirement from 1 to 6, the proportions of the strategies show 90% in bin number 1, the other 10% on the right side (Figure 4a).

Cluster 2: (27 histograms) We notice that this cluster also has two groups. The first group occurs if visibility values go from 0 to 0.3, obstinacy 4 to 6, and cooperation demand is 1 or 2, the strategies in bin 1 are established from 70% to 80% the rest of the clusters are on the left side up to bin -4. The second group occurs when visibility is from 4 to 7, obstinacy from 2 to 5, and the cooperation demand goes from 1 to 3, around 10% groups of agent strategies are established in bin -2 to bin -4, the rest are in bins -1, 0 and 1 (Figure 4b).

Cluster 3: (63 histograms) With visibility conditions from 0 to 0.5, obstinacy from 2 to 6, and demand for cooperation from 4 to 6, it turns out that the discriminating agents predominate. In addition, there are groups in which, with some symmetry, they are cooperators and defectors. There are no severe defectors (Figure 4c).

Cluster 4: (144 histograms) In this cluster, the visibility is between 0.1 to 0.8, obstinacy goes 3 to 6, and cooperation demand from 1 to 6, there is a concentration of 40% to 50% of discriminators, and the remaining percentage are cooperators, including some unconditional cooperators. There are very few defectors, or in some cases, they do not appear. (Figure 4d).

Cluster 5: (20 histograms) This cluster shows that when visibility varies from 0 to 0.6, obstinacy from 0 to 1, and cooperation demand from 5 and 6, total and partial polarization is generated. They are generally 50% and 50% or 60% and 40% altruists and defectors. In complete polarization, there are only unconditional cooperators and severe defectors (Figure 4e).

Cluster 6: (56 histograms) We observe that this cluster has two groups, one with five histograms of heterogeneous form, which occur when there is visibility 0, obstinacy from 2 to 6, and cooperation demand is 6. The other group has a "Semi-heterogeneous" form; the cooperators and defectors are in percentages of 30% to 70%. The visibility varies from 0.4 to 0.9, the obstinacy from 0 and 1, the cooperation demand varies from 1 to 5 (Figure 4f).

Cluster 7: (16 histograms) If the visibility parameters vary from 0.2 to 0.4, obstinacy from 0 to 1, and cooperation demand from 2 to 5, then groups are formed where the majority are defectors, 80% to 90%, and the others are cooperators and discriminators (Figure 4g).

Cluster 8: (25 histograms) Visibility and obstinacy are low, with values 0 to 0.3 and 0 to 1, respectively, and cooperation demand from 1 to 5, the groups of strategies are formed on the right side, and in very few there are on the left side of the histogram. In other words, a society of defectors is established (Figure 4h).

Cluster 9: (166 histograms) We have a society of discriminatory agents if the cooperation demand is null or the visibility is total (Figure 4i).

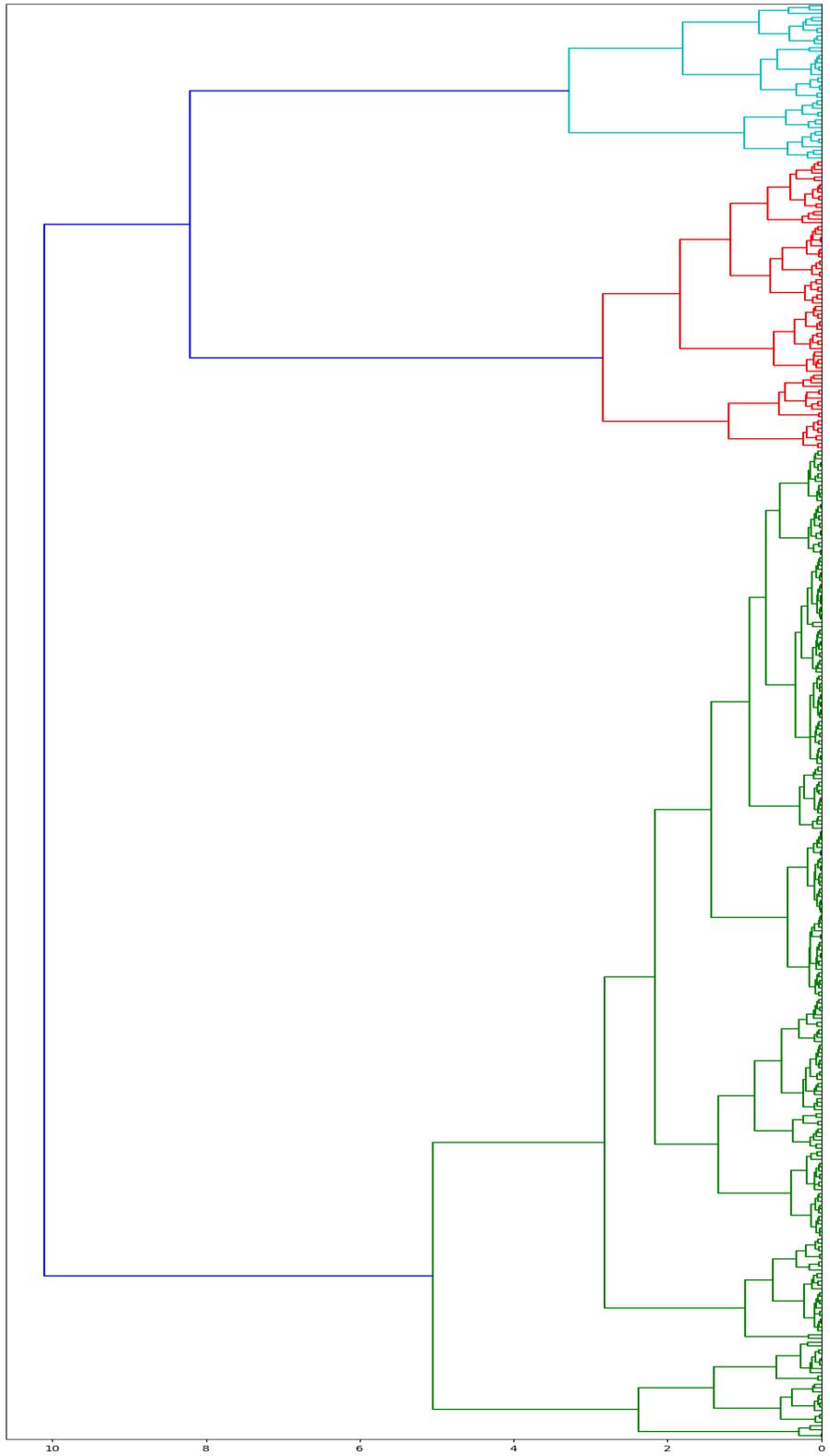

Figure 5.

## Discussion

In this work, the HCA was used to group the 539 simulations. We analyzed the histograms based on the characteristics space to discriminate or group them based on similarity. We used the final value of the simulation and the triple of the initial conditions (the parameter space) to perform the analysis of the simulations.

The histograms obtained from each simulation were analyzed using HCA and K-means. The K-means, a multivariate analysis method, was applied to discriminate different characters based on different parameters. However, the K-means did not reveal the precise relationships between the different simulations and changed each time it was run. The dendrograms obtained by HCA intuitively represented the relationships between the simulations in terms of mean, standard deviation, and skewness. In the HCA, the samples obtained were passed into the parameter space and then represented by a Convex hull (clusters) in a three-dimensional space (Figure 4).

Cluster 1. In the agglomeration of histograms grouped in group 1 of cluster 1, there is null visibility; none of the agents can observe the reputation value of the others. These conditions cause 80% of agents grouped as discriminators, and another 20% of the agents are distributed around as cooperators and defectors; this is because the agents are obstinate to change their strategy and are demanded to cooperate a lot (when only the demand for cooperation is reduced, they become more cooperative, see cluster 3). The second group of cluster 1, arises when the visibility goes from 70% to 80%, they are required to cooperate a lot, and these agents tend to change their strategy easily; the discriminating agents tend to be 90%, and the remaining 10% of the agents are on the right side of the histograms, being these defectors.

Cluster 2. Being very obstinate to change their strategy, demanding little cooperation between them and since they have 20% visibility to see the reputation, there are 90% of discriminatory agents and 10% of agents are cooperators, since that with that percentage they are trusting that they will have to cooperate with agents who cooperate. However, when visibility increase over 50%, the agents tend to be less discriminatory, agglomerating in 80% of agents and the other 20% of the agents are cooperators, considering that their indicator to change its strategy is 3 to 4 and that they are required to cooperate at a minimum.

Cluster 3. If an agent hardly knows the reputation of others, he has high obstinacy, and he is demanding a lot to cooperate; there are 40% of discriminators and, there are both cooperators and defectors. Partnerships are created with cooperators that have some unconditional cooperators. Unconditional cooperators are those who help everyone else regardless of reputation. Defector societies also emerge. Societies emerge with a certain symmetry of cooperators and discriminators.

Cluster 4. It is the second cluster with more histograms; it has a total of 144. In this, societies were grouped in proportions of 60-40. The 60% grouping is built of discriminators, and the 40% is built of cooperators. This situation happens with average visibility between 30% to 60%, an average obstinacy of 80%. The cooperative society tends to coexist with the discriminators, perhaps because, at some point, an agent will meet another who will pay them back.

Cluster 5. When the environment has the conditions that agents know less than 60% of people's reputations, there is a high cooperation demand, and agents can be quickly readapted; a polarized society is created, that is, society is divided into two groups: those who cooperate a lot and those that do not cooperate at all. In this case, if an agent is a severe defector, he can take advantage of the high demand for cooperation.

Cluster 6. Two types of histograms were grouped in this cluster. The first society type is made up where reputation is known at 0%, and there is a 100% requirement for cooperation. In this case, there is a heterogeneously distributed society. The second society has around 30% cooperators and 70% defectors, with the conditions of average visibility of 50%, obstinacy, and a requirement to cooperate at 30%. With these parameters and the principles of the model, it happens that there are agents always willing to cooperate with others, even if they get little pay or nothing.

Cluster 7. This cluster shows a society with 80% of discriminators. However, there are a few cooperators and super cooperators, and this happens when the reputation is only known in no more than 40%, there is an intermediate obstinacy, and they are required to cooperate in general.

Cluster 8. There are approximately 30% discriminators and 70% defectors; this happens when agents in society are obstinate, have little visibility of reputation, and the demand for cooperation is regular or null. In these conditions, defectors reinforce their condition since average obstinacy and good conditions of their cooperation mean that as soon as the benefits are obtained, they return to their position of defectors. The defectors in these simulations can become severe defectors.

Cluster 9. In this cluster, societies are made up entirely of discriminators. It happens in one or other of the two initial conditions: there is no cooperation demand, or everyone can always see the reputation of others. As the main characteristic of the model is individualism, then since there is no cooperation demand or always knowing the reputation of others, the position of helping the person who has helped is fixed (although limited), so the discriminators are established.

There are four clusters with a whole domain of discriminators (cluster 1, 2, 4, and 9); 359 of the 539 simulation experiments, 66% of the total cases. If visibility, obstinacy, and demand for cooperation are low, communities of discriminators are established (cluster 2). Furthermore, when we have the intermediate conditions of invisibility, obstinacy, and demand for cooperation (cluster 4), the vast majority of the population tends to be discriminatory (except for a few defectors). For extreme conditions of null (or almost null) values invisibility, demand for cooperation, and obstinacy, there are situations where discriminatory or defectors predominates (clusters 8 and 9).

A defector society is established (more than 80% of this agent type) when visibility and stubbornness are low, and the demand for cooperation varies from 1 to 5 (Cluster 8). However, there is no demand for maximum cooperation as it goes through polarization (cluster 5), going through an intermediate process (cluster 7).

There are practically no discriminators when visibility does not exceed 0.5, obstinacy is less than 3, and the cooperation demand is greater than 1 (for clusters 5, 7, and 8). There are no cooperators when we have high visibility or no cooperation requirement (cluster 9), but also cooperators disappear with little visibility, little obstinacy, and intermediate demand for cooperation (cluster 1 and 8).

It seems that the conditions for not appearing defectors, or at most 10%, are in the center of the parameter space (joining the convex-hulls of clusters 2 and 4). That is, in the intermediate conditions of the parameter space of our model, almost no defectors arise, but this also happens with null conditions of cooperation demand or total visibility (cluster 9)

If we consider the initial conditions as the environment and we mark as special situations when these take the minimum or maximum values, we have that in the neutral environment (all parameters in 0: $V = 0$, $CD = 0$ and $T = 0$), we have an environment of discrimination, as well as in extreme environments ($V = 1$, $T = 6$ and $CD = 0$; $V = 1$, $T = 0$, $CD = 6$ and $V = 1$, $T = 0$, $CD = 6$). In the extreme environment (some of the parameters at the limit: $V = 0$, $CD = 6$ and $T = 6$), we have a heterogeneous society.

In the conditions of polarization that occur when there is a high demand for cooperation, little visibility, and obstinacy, it can be interpreted that the defectors demand more and more, and the cooperators are always required to cooperate. The characteristic of individualism makes these values of the parameters benefit from being defectors (cluster 5).

Finally, when there are more cooperators and discriminators in the histogram, the cluster is large and reciprocally.

## Conclusions

This work, a new simulation model (based on ABM) is proposed, called IIRM, based on the indirect reciprocity mechanism by reputation, to analyze the cooperation behavior with individualism.

This model includes the cooperation demand that can be interpreted as a social norm, conditions of the social substrate such as visibility (of reputation) and, agents may have the obstinacy phenotype.

The use of data science methods was a helpful tool to determine and visualize the behavior patterns of the multiple results of the simulations; the results are efficiently displayed in different clusters, where each one of them has similar behaviors according to the strategy, this thanks to hierarchical clustering method. The HCM shows that there are very close clusters (in the parameter space) because by varying some of the parameter values a little, one passes from one cluster to another.

In the IIRM, we generate multiple simulation results; this research covers a complete category of simulation experiments taking all the parameter values.

In general, the results show that agents tend to be individualistic and have a discriminatory stance. Only under exceptional conditions does one have an environment in which there is cooperation. The simulation results did not show scenarios in which most of the agents are cooperators, regardless of how obstinate they are and forced to cooperate.

Due to the simulation results in the IIRM with the heterogeneous initial conditions (of our model), it is impossible to have a society of pure cooperators.

There are cases where there are only unconditional cooperators and severe defectors; when one of these groups appears, the other appears, which makes us think that they are in a certain symbiosis. The simulation results also show that discriminating societies can coexist with defectors and cooperators.

Our results show that some simulation cases establish heterogeneous societies. Some results also show that a pure defectors society can be established when there is very little information of the agents' reputation (cluster 8).

From the simulation results in the IIRM, it could be said that there are no conditions for cooperation to be established (most of the population is cooperative), although there are conditions to arise the cooperation (cooperators appear). Although it seems beneficial that there are no discriminators, they are necessary because if not, increase severe defectors, they use cooperators; the conditions in which they arise are extreme visibility, minimal obstinacy, and demand for maximum cooperation.

Unlike the other models in the literature, this model considers internal attributes such as obstinacy, external characteristics such as cooperation demand, and the one that has been used in other investigations, probability of seeing reputation.

This model has two new parameters in the experiments and coincides with the statement of Nowak and Sigmund (Nowak & Sigmund, 1998a); altruism favors cooperation but with special conditions of the parameters.

Our work is different from the article (Nowak & Sigmund, 1998b) because we do not consider the proportions of three agents' types and different forms of payments and criteria of p, to know if an agent is G-individual. It also differs from the work of (Lotem et al., 1999), since they are interested in identifying the effect of a people with a D = + 7 phenotype, fixed by all simulation generations, which we do not do. Leimar and Hammerstein in (Leimar & Hammerstein, 2001), analyze standing strategies or good standing to compare with image scoring, which also differs from ours.

Research that also (Manfred Milinski et al., 2001) did, in a real experiment and with a phenotype player called "no player". It is also different from the work of (Mohtashemi & Mui, 2003) since they analyze the evolution of indirect reciprocity through social information and determine this evolution with simulations and analytical calculations.

Since our model promotes individuality and includes social parameters, our results are not quite like previous papers. In a certain sense, we agree with Suzuki (Suzuki & Kimura, 2013) since the parameter of cooperation demand is like a social norm.

Although our work is different from the work of (Seinen & Schram, 2006) due to the experimental evidence, we have that the concepts of the "groups of social norms" have similarity to the variation of the cooperation demand parameter of our model.

# References

Alexander, R. D. (1979). Natural Selection and the Analysis of Human Sociality. In *Changing Scenes in the Natural Sciences, 1776-1976* (Vol. 12, pp. 283–337). Philadelphia Academy of Natural Sciences.

Alexander, R. D. (1981). *Darwinism and Human Affairs - Richard D. Alexander - Google Libros*. Univ.of Washington Press.

Alexander, R. D. (1987). *The biology of moral systems*. (Routledge, Ed.). New York, NY, USA.

Baddeley, A. (1987). Working memory, Oxford 1986. *Applied Cognitive Psychology*, *2*(2), 166–168. https://doi.org/10.1002/acp.2350020209

Boerlijst, M. C., Nowak, M. A., & Sigmund, K. (1997). The logic of contrition. *Journal of Theoretical Biology*, *185*(3), 281–293. https://doi.org/10.1006/jtbi.1996.0326

Collins, A., Petty, M., Vernon-Bido, D., & Sherfey, S. (2015). A call to arms: Standards for agent-based modeling and simulation. *JASSS*, *18*(3). https://doi.org/10.18564/jasss.2838

Gonzalez-Silva, M. I. (2018). *Modelación y simulación de un nuevo modelo del mecanismo de cooperación de reciprocidad indirecta*. Tesis MenC., Universidad de Guadalajara.

González Silva, R. A., González Silva, M. ., & Juárez Lopéz, H. A. (2020). Modeling and simulation of the phenomenon of cooperation in the mechanism of Individualistic Indirect Reciprocity. In A. Aguilera Ontiveros & N. L. Abrica Jacinto (Eds.), *Generating Knowledge Through Modeling : An Exploration of Computational Sociology* (In process, p. 300). México.

Grimm, V., Berger, U., Bastiansen, F., Eliassen, S., Ginot, V., Giske, J., … DeAngelis, D. L. (2006). A standard protocol for describing individual-based and agent-based models. *Ecological Modelling*, *198*(1–2), 115–126.

Grimm, V., Berger, U., DeAngelis, D. L., Polhill, J. G., Giske, J., & Railsback, S. F. (2010). The ODD protocol: A review and first update. *Ecological Modelling*, *221*(23), 2760–2768. https://doi.org/10.1016/j.ecolmodel.2010.08.019

Grimm, V., Railsback, S. F., Vincenot, C. E., Berger, U., Gallagher, C., DeAngelis, D. L., … Ayllón, D. (2020). The ODD protocol for describing agent-based and other simulation models: A second update to improve clarity, replication, and structural realism. *JASSS*, *23*(2). https://doi.org/10.18564/jasss.4259

Hamilton, W. D. (1964). The genetical evolution of social behaviour. I. *Journal of Theoretical Biology*, *7*(1), 1–16.

Hennig, C., Meila, M., Murtagh, F., & Rocci, R. (2016). Handbook of cluster analysis.

J. Gary Polhill, D. P. D. B. and V. G. (2008). Using the ODD Protocol for Describing Three Agent-Based Social Simulation Models of Land-Use Change.

King, R. S. (2015). *Cluster Analysis and Data Mining. An Introduction*.

Laatabi, A., Marilleau, N., & Nguyen-huu, T. (2018). ODD + 2D : An ODD based protocol for mapping data to empirical ABMs, (March). https://doi.org/10.18564/jasss.3646

Leimar, O., & Hammerstein, P. (2001, April 7). Evolution of cooperation through indirect reciprocity. *Proceedings of the Royal Society B: Biological Sciences*. Royal Society. https://doi.org/10.1098/rspb.2000.1573

Lotem, A., Fishman, M. A., & Stone, L. (1999). Evolution of cooperation between individuals. *Nature*, *400*(6741), 226–227. https://doi.org/10.1038/22247

Milinski, M, & Wedekind, C. (1998). Working memory constrains human cooperation in the Prisoner's Dilemma. *Proceedings of the National Academy of Sciences of the United States of America*, *95*(23), 13755–13758. https://doi.org/10.1073/pnas.95.23.13755

Milinski, Manfred, Semmann, D., Bakker, T. C. M., & Krambeck, H. J. (2001). Cooperation through indirect reciprocity: Image scoring or standing strategy? *Proceedings of the Royal Society B: Biological Sciences*, *268*(1484), 2495–2501. https://doi.org/10.1098/rspb.2001.1809

Mohtashemi, M., & Mui, L. (2003). Evolution of indirect reciprocity by social information: The role of trust and reputation in evolution of altruism. *Journal of Theoretical Biology*, *223*(4), 523–531. https://doi.org/10.1016/S0022-5193(03)00143-7

Nowak, M. A. (2006). *Evolutionary dynamics : Exploring the equations of life*. Cambridge, Massachusetts, USA.

Nowak, M. A., & Sigmund, K. (1998a). Evolution of indirect reciprocity by image scoring. *Nature*, *393*(6685), 573–577. https://doi.org/10.1038/31225

Nowak, M. A., & Sigmund, K. (1998b). *The Dynamics of Indirect Reciprocity*. J. theor. Biol (Vol. 194).

Nowak, M. A., & Sigmund, K. (1998c). The Dynamics of Indirect Reciprocity. *Journal of Theoretical Biology.*, *194*, 561–574.

Nowak, M. A., & Sigmund, K. (2005, October 27). Evolution of indirect reciprocity. *Nature*. Nature Publishing Group. https://doi.org/10.1038/nature04131

Ohtsuki, H., & Iwasa, Y. (2006). The leading eight: Social norms that can maintain cooperation by indirect reciprocity. *Journal of Theoretical Biology*, *239*(4), 435–444. https://doi.org/10.1016/j.jtbi.2005.08.008

Okada, I. (2020). A review of theoretical studies on indirect reciprocity. *Games*, *11*(3), 1–17. https://doi.org/10.3390/g11030027

Paranjape, R., Wang, Z. G., & Gill, S. (2018). *Agent-Based Modeling and Simulation*. *Intelligent Systems Reference Library* (Vol. 133). https://doi.org/10.1007/978-3-662-56291-8_2

Rand, D. G., & Nowak, M. A. (2013). Human cooperation. *Trends in Cognitive Sciences*, *17*(8), 413–425.

Rockafellar, R. T. (1990). *Convex Analysis*. *Princeton University Press*. Retrieved from https://www.jstor.org/stable/j.ctt14bs1ff

Seinen, I., & Schram, A. (2006). Social status and group norms: Indirect reciprocity in a repeated helping experiment. *European Economic Review*, *50*(3), 581–602. https://doi.org/10.1016/j.euroecorev.2004.10.005

Suzuki, S., & Kimura, H. (2013). Indirect reciprocity is sensitive to costs of information transfer. *Scientific Reports*, *3*, 1–5.

Van Rossum, G., & Drake, F. L. (2009). Python 3 Reference Manual. Scotts Valley, CA: CreateSpace. Retrieved January 28, 2021, from http://citebay.com/how-to-cite/python/

Wilensky, U. (1999). NetLogo, Center for Connected Learning and Computer-Based Modeling, Northwestern University, Evanston, IL. Retrieved from http://ccl.northwestern.edu/netlogo/

Wrigth, S. (1943). Isolation by distance. *Genetics*, *28*(march), 114–138.

# Appendix

**The ODD protocol of MC-RII**

**Overview**

The model we present is a variant of the Nowak and Sigmund model (Nowak & Sigmund, 1998a), based on agents and developed with the NetLogo 6.1.1 program. This simulation software provides a comprehensive modeling environment based on its agent-oriented language.

1. Purpose.

Model and simulate the evolution of cooperation through the individualistic indirect reciprocity cooperation mechanism in artificial societies. Agents are characterized by reinforcing their strategy when they obtain good results (payments). In the agents' environment, the ability to see the reputation of others, the degree of obligation to cooperate, and their obstinacy to change strategy is handled.

2. Entities, state variables and scales.

A population of n agents is established; each one has three attributes or variables, a strategy k with a range of integer values from -5 to 6, a reputation s with integer values from -5 to 5, and a payment function p, bounded with values from -5 to 5. The global variables are visibility (V), obstinacy (O) and cooperation demand (CD), with values at $\{0, 0.1, 0.2, \cdots, 0.9, 1\}$, $\{0, 1, \cdots, 6\}$ and $\{0, 1, \cdots, 6\}$, respectively. The time scale is in discrete steps.

3. General description of the process and planning.

The simulation model is represented by a population of n = 100 agents, the players have a reputation and payout function with initial value s = 0 and p = 0, respectively. In each experiment, the initial values of visibility (V), obstinacy (O), and cooperation demand (CD) are established. Each experiment consists of 200-time steps.

At each time step, a generation is created; in each generation m (interactions) pairs of agents are randomly chosen, one as possible donor i and the other as recipient j. The agents know with some probability V, the reputation of the agents. With this, there is a dynamic stochastic and unidirectional interaction.

At each time step in the interaction dynamics, the donor agent i must decide to cooperate or not with recipient agent j. A donor i cooperates if the value of its strategy $k_i$ is less than or equal to the reputation $s_j$ of recipient j, that is, $k_i \leq s_j$. When donor i cooperates, his reputation score increases by one unit, and if he doesn't, it decreases by one. When cooperation occurs, the donor pays a cost, c, and the recipient obtains a benefit, b. There is no reward in the absence of cooperation.

Each player has, on average, 2m / n interactions. The attributes updating dynamics is asynchronous; the agent variables values update is according to its interaction.

**Pseudo-code of the evolutionary dynamics of the simulation model**

1  Initialize:
   t = 200, N = 100, k = {5-, -4, ... 5, 6}, s = 0, p = 0, O = (0, 6), v = (0, 1), CD = (0, 6), c = 0.1, b = 1
2      for each round do
3          round t
4          for each agent do
5              randomly donor/receptor
6              if visibility > random(0, 1) then
7                  if $k_i \leq s_j$, then
8                      cooperative action for donor and receptor
9                      update obstinacy
10                 else
11                     action of defecting for donor and receptor
12                     update obstinacy
13                 end if
14             else
15                 if $k_i$ < cooperation demand, then
16                     cooperative action for donor and receptor
17                     update obstinacy
18                 else
19                     action of defecting for donor and receptor
20                     update obstinacy
21                 end if
22             end if
23             update (k,s,p)
24         end for
25     return (k, s, p) for next round
26     output(k, s, p)
27     end for

**Design Concepts**

4. Design of Concepts

The interaction dynamics rules used in the simulation models are designed to represent the individual and group behavior of the agents.

- Basic principles. The individualistic indirect reciprocity mechanism is based on Alexander's indirect reciprocity for reward principle (Alexander, 1979, 1981).
- Emergency. This is shown with the behavior patterns of the frequency distribution of k cooperation. Groups of cooperators, discriminators, and defectors emerge in the final simulations; polarized, heterogeneous, and centrist societies.
- Goals. Agents have the option of cooperating or defect. They do this based on a comparison between their strategy and the recipient's reputation score.
- Learning. Each agent strengthens her strategy based on the benefits generated in the previous iteration.

- Prediction. The reputation score provides information on whether the receiving agent is a good candidate for cooperation.
- Detection. Donors perceive the reputation score from the recipients.
- Interaction. In each round, two randomly selected agents interact; one is a potential donor, and the other is a potential recipient.
- Randomness. Two agents are randomly selected from the entire population in each round.
- Observation. The evolution of cooperation is observed through the strategy values' frequency distribution.

**Details**

5. Initialization.

It begins with a population of n = 100 agents. The agents' reputation score is set at s = 0, and strategy k is homogeneously distributed with a range of -5 to 6.

6. Entrance.

The values of the parameters will vary in each simulation experiment: V∈ {0,0.1,0.2, ⋯, 0.9,1}, O∈ {0,1, ⋯, 6} and CD∈ {0,1, ⋯, 6}. The parameter values: b = 1, c = 0.1 are fixed, to avoid negative payments we add 0.1 at the beginning of each interaction.

7. Sub-models

Each agent has a strategy $k$, image scoring $s$ and payment function $p$. The model parameters are visibility ($V$), obstinacy ($O$) and cooperation demand ($CD$). If agent $i$ is a donor and agent $j$ is a receptor, the actualization agents' values are defined as follows.

| | Table $k_j \leq s_i$ | |
|---|---|---|
| | **Agent i** | **Agent j** |
| $k_*(t+1) =$ | $k_i(t)$ | $k_j(t) + 1$ si $k_j(t) > 0$ <br> $k_j(t) - 1$ si $k_j(t) < 0$ |
| $s_*(t+1) =$ | $s_i(t) + 1$ | $s_j(t)$ |
| $p_*(t+1)=$ | $p_i(t) - c$ | $p_j(t) + b$ |

| | Table $k_j \geq s_i$ | |
|---|---|---|
| | **Agent i** | **Agent j** |
| $k_*(t+1) =$ | $k_i(t)$ | $k_j(t) - 1$ si $k_j(t) > 0$ <br> $k_j(t) + 1$ si $k_j(t) < 0$ |

| $s_*(t+1) =$ | $s_i(t) - 1$ | $s_j(t)$ |
|---|---|---|
| $p_*(t+1)=$ | $p_i(t)$ | $p_j(t)$ |